\documentclass[letterpaper, 10 pt, conference]{ieeeconf}  

\IEEEoverridecommandlockouts                              

\usepackage{graphicx} 
\usepackage{amsmath} 
\usepackage{amssymb}  
\usepackage{subcaption}

\newcommand{\matr}[1]{\mathbf{#1}}
\usepackage{xcolor}

\title{\LARGE \bf
Data-driven Spatio-Temporal Scaling of Travel Times for AMoD Simulations
}

\author{Arslan Ali Syed$^{1}$, Yunfei Zhang$^{1}$ and Klaus Bogenberger$^{1}$
\thanks{$^{1}$All authors are with the Chair of Traffic Engineering and Control, Department of Mobility Systems Engineering,
        Technical University of Munich, 80333 Munich, Germany
        {\tt\small arslan.syed@tum.de}}%
}

\begin{document}

\maketitle
\thispagestyle{empty}
\pagestyle{empty}

\begin{abstract}


With the widespread adoption of mobility-on-demand (MoD) services and the advancements in autonomous vehicle (AV) technology, the research interest into the AVs based MoD (AMoD) services has grown immensely. Often agent-based simulation frameworks are used to study the AMoD services using the trip data of current Taxi or MoD services. For reliable results of AMoD simulations, a realistic city network and travel times play a crucial part. However, many times the researchers do not have access to the actual network state corresponding to the trip data used for AMoD simulations reducing the reliability of results. Therefore, this paper introduces a spatio-temporal optimization strategy for scaling the link-level network travel times using the simulated trip data without additional data sources on the network state. The method is tested on the widely used New York City (NYC) Taxi data and shows that the travel times produced using the scaled network are very close to the recorded travel times in the original data. Additionally, the paper studies the performance differences of AMoD simulation when the scaled network is used. The results indicate that realistic travel times can significantly impact AMoD simulation outcomes.

\end{abstract}




\section{Introduction}\label{sec:intro}

The last decade saw the widespread emergence of Mobility-on-Demand (MoD) services. Using smartphone applications, the customers in these services can seamlessly request rides from origin to destination. With advancements in autonomous vehicle (AV) technology, the AVs based MoD (AMoD) services are also on the horizon.

Seeing the potential arrival of AMoD services, there has been a steady increase in research on the potential benefits of AVs in MoD services. A major advantage of AMoD services is the central control of the whole fleet which leaves the allocation of customers to vehicles entirely on the central fleet controller (FC). Many researchers have used agent-based simulations to evaluate the efficiency of different FCs. These simulation studies often use the customer trips data of the current MoD or Taxi services to study the scenario where the fleet consists of only AVs. They report a significant performance improvement by introducing AVs instead of human drivers \cite{hyland2018dynamic, dandl2021regulating}. 

However, the agent-based simulation must be as realistic as possible for these outcomes to be reliable. In this regard, the accuracy of the network travel times plays a significant role. Since the actual state of the city network corresponding to the customer data used is usually unavailable, researchers use different techniques to replicate realistic travel times. To have a better overview of the literature, the following compares various AMoD studies from the perspective of travel time estimation. They can be mainly divided into three categories:

\begin{itemize}
    \item Fixed travel time: This assumes that the travel time does not change over time. This includes directly using the free-flow speeds from available maps data like OpenStreetMap (OSM) \cite{syedAsynchronousAdaptiveLarge2019, zhangSimulatingChargingProcesses2022} or multiplying the whole network by a common factor to replicate different hours of the day, such as peak and off-peak hours \cite{chen2016operations} or average travel speed~\cite{hyland2018dynamic, hyland2020operational, alonso-moraOndemandHighcapacityRidesharing2017}.
    
    \item Simulation-based methods: These studies directly use the microsimulators for AMoD simulation, such as Aimsun \cite{dandl2017microsimulation}, or use their outputs as a traffic state estimation to be used in a simpler agent-based simulation ~\cite{dandl2021regulating, engelhardtQuantifyingBenefitsAutonomous2019}. The microsimulation is first calibrated for realistic travel times using a third data source such as the data of loop detectors. Usually, the calibration data is not easily available and does not necessarily correspond to the time and day of the customer data simulated.
    
    \item Data-driven methods: These studies utilize historical data to estimate time-dependent scaling factors without a microsimulator. They can be based on separate real data of past trips in the city \cite{jager2018multi} or using average scaling factors derived from the same customer trips as used for AMoD simulation \cite{syedDensityBasedDistribution2021}. Some studies also derive zone-specific factors derived from MFD estimated by loop detector data~\cite{zhang2023}.
\end{itemize}

However, these three methods have their flaws: fixed travel time is not able to capture the temporal dynamics, simulation-based methods are computationally expensive, and researchers often do not have access to the data required for calibration, and data-driven approaches often do not consider the changing link-level travel times. Furthermore, since many studies do not aim to reproduce the travel times of the original trips, they can lead to a significant deviation from the trip travel times recorded in the historical customer data, which can ultimately impact the AMoD simulation results.

In contrast to the above, the current research suggests that a significantly more realistic simulation can be produced by spatio-temporal scaling of the network travel times using customer data. 
Using customer data is not a new idea: \cite{santi2014quantifying, ghandeharioun2023real} have used the customer data (OD and trip duration) to estimate the travel time over a period using heuristics by segmenting the whole path into roads between intersections.
Unlike these methods, the paper first divides the historical data into fixed periods. Then it solves an optimization problem for each period such that most customers have the exact travel times between the origin and the destination as recorded in the data set. Thus, each customer trip serves as a sample of the network state at that specific time of the day. The aim is to produce a network state similar to when the data was recorded.

\section{Methodology}

In literature, the performance of AMoD services is often evaluated in agent-based simulation. The customer requests are either generated using artificial data or using real trip data of existing alternatives to the AMoD services like taxi services or MoD services. The current paper assumes the latter as it is not possible to derive realistic network travel times using artificial trips.

This paper argues that the same data used for simulating the AMoD services can also be used to scale the link-level network travel times. This improves the realism of the entire simulation and makes the results more reliable. Typically, the data providers of real trips use various levels of spatio-temporal aggregation to make the data anonymous before its publication. The technique used in the paper assumes a moderate level of data anonymity where the data set at least provide the following information: 1) the exact coordinates of the trip origin and destination locations 2) the total trip time and 3) the total trip distance. However, the actual trip trajectory is removed from the data. 

The fundamental concept here is that for reliable AMoD simulation results, the network travel times must be scaled such that any trip on the scaled network have a similar travel time as recorded in the original data. Here the network used for the AMoD simulation is assumed to be rather simpler: all lanes on the link between two nodes are assumed to have the same speed and the vehicles traverse the edges independent of the other vehicles on the link. However, the network is still assumed to consist of a directed graph, and thus, each link can have different speeds in opposite directions.

The following sections describe an often-used scaling approach used in literature for the AMoD simulation followed by the spatio-temporal scaling method presented in the current paper.



\subsection{Mean Factor Method (MFM)}
The MFM scales the network travel times dynamically according to the historical trip data. However, it only considers the temporal aspect for this purpose. The historical trip data are divided into periods of $\Delta T_{scale}$ using the trip start times. Then, for a single temporal group, the MFM scales all of the edges by a common scaling factor~\cite{syedDensityBasedDistribution2021, dandl2020network}. It first calculates the sum of the travel times of all historical trips within a single $\Delta T_{scale}$ period, represented by $t^{data}_{od}$. Then the sum of the travel times for all trips is calculated using the free-flow speeds of the network, represented by $t^{flow}_{od}$. The travel times of all the edges of the network are then multiplied by the mean travel time factor:
\begin{equation}
    f_{mean} = \frac{t^{data}_{od}}{t^{flow}_{od}}
\end{equation} 

MFM has the advantage that a single travel time matrix can be used for the entire simulation; the matrix is loaded at the beginning of the simulation and multiplied by the corresponding $f_{mean}$ for the current simulation time. However, since it uses the sum of travel times for scaling, the operational regions with the highest number of trips can weigh significantly higher than those in other regions. Thus, it represents the actual travel times in high-demand areas much better than in lower-demand areas. Additionally, it fails to capture the speed variation based on spatial features.

\subsection{Spatio-Temporal Scaling Method}

\begin{figure}[tbp]
  \centering
    \includegraphics[width=0.8\linewidth]{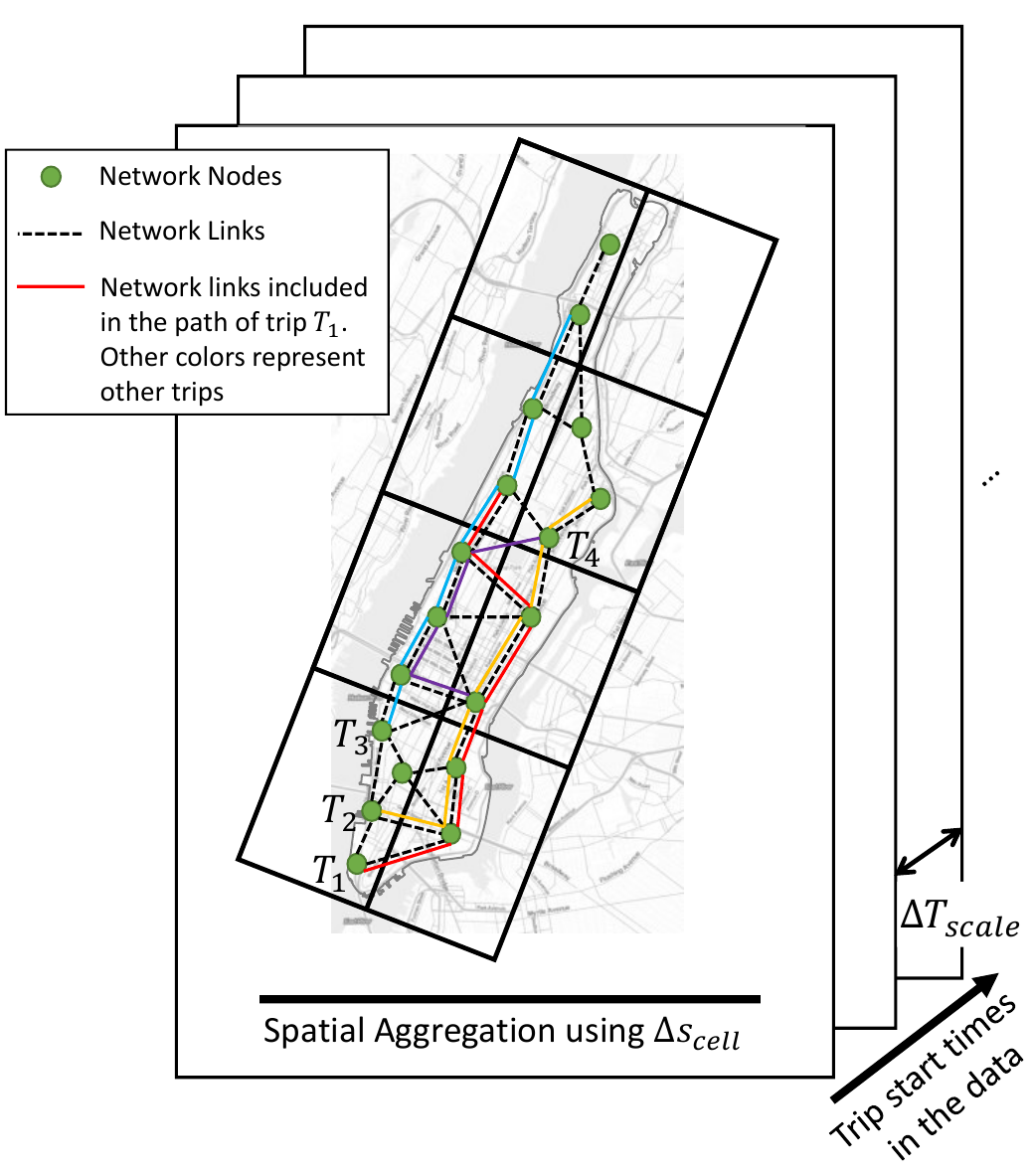}
  \caption{An example of spatio-temporal grouping of customer data for scaling the city network. For simpler visualization, the directional edges between the nodes are reduced to a single edge.}
  \label{fig: network example}
\end{figure}

In contrast to the MFM method which only uses temporal information for the scaling of travel times, the spatio-temporal scaling method additionally uses spatial information for this purpose. This allows a more accurate link-specific scaling of the network by solving an optimization problem for each $\Delta T_{scale}$. It is fundamentally based on the concept that each historic trip provides a snapshot of the network state at that particular time of the day. If multiple trips are grouped together using temporal information, i.e., a period of $\Delta T_{scale}$, these trips can potentially provide multiple observations of the overlapping links on their paths from origin to destination. The paper develops two variants of the scaling method based on the objective function used for optimization.

The first method is referred to as the squared scaling method (SSM). It divides the whole operational area into regions $Z_{scale}$ using a regular grid of cell size $\Delta s_{cell}$. Let $E$ represent the edges of the network used for AMoD simulation, $t_e^{flow}$ represent the free-flow travel time of an edge $e \in E$ and $E_z \subseteq E$ represent the edges that are entirely within region $z \in Z_{scale}$. Some of the edges may be located in multiple regions, denoted by $E_{m} \subseteq E$. For each of these edges $e_{\hat{z}} \in E_m$, the set $\hat{z} \subseteq Z_{scale}$ represents the regions the edge belongs to. Afterward, it calculates the shortest paths for each of the trips within the period $\Delta T_{scale}$. Let the set $C$ represent all the trips within the period $\Delta T_{scale}$, then the shortest path of a trip $c \in C$ consists of a sequence of network edges, represented by $E_c \subseteq E$. The main purpose of the method is to scale the free-flow travel time of each edge in $E_c$ for all $c \in C$ such that the travel times comes as close as possible to the historical travel time $t_{od}^c$ for $c \in C$. If two or more trips in $c \in C$ have the same origins and destinations, then the mean of the travel times of these trips is used. 

Let $E_u \subseteq E$ represent the union of all edges in the shortest paths of $C$, i.e., the union of $E_c$ for all $c \in C$. Then, as the first step, the method only scales the edges in $E_u$ by solving the following optimization problem.

\begin{subequations}
    \label{eq: squared scaling}
    \begin{align}
        ~& \min_{\matr{x}} \sum_{c \in C} \bigg( t_{od}^c - \underbrace{\sum_{e \in E_c} x_e t_e^{flow}}_{\text{scaled travel time}} \bigg)^2 \label{eq: squared scaling objective} & ~ \\
        \textrm{s.t.} & \quad x_e t_e^{flow} \leq t_e^{max}         & \forall e \in E_u \\
        ~& \quad x_e \geq 1         & \forall e \in E_u \\
        ~& \quad \matr{x} \in \mathbb{R}^{\vert E_u \vert} \nonumber 
    \end{align}
\end{subequations}
where $\matr{x}$ is a vector of linear variables used for scaling each edge within $E_u$ and $t_e^{max}$ is the maximum allowed travel time on the edge. $t_e^{max}$ is calculated using the edge length ($d_e$) and a parameter for the minimum travel speed allowed on the edge ($S_e^{min}$), i.e., $t_e^{max} = d_e / S_e^{min}$.

After calculating the scaling factors for the edges in $E_u$, the SSM scales the rest of the edges in $E$. These additional edges are divided into two categories. First, the ones belonging to a single region (i.e., belonging to $E_z$) are scaled using the mean of the scaling factors already calculated in the first step for that region, i.e., they are scaled by the mean of all $x_e \in E_u \cap E_z$. Second, the edges that fall within multiple regions are scaled using the mean of the scaling factors for all edges within those zones, i.e., for an edge $e_{\hat{z}} \in E_{m}$ it is scaled by the mean of all $x_e \in \cup_{z \in \hat{z}} (E_u \cap E_z)$. The main assumption is that the scaling factor should be on a similar scale in the immediate surrounding of the scaled edges $E_u$. If any edge is still not scaled using these two approaches, then they are scaled using the mean of $\matr{x}$. 

The second spatio-temporal scaling method uses the same approach as SSM; however, instead of using squared error for the objective function, it uses absolute error. Thus, the method is referred to as the absolute scaling method (ASM) and Eq.~\ref{eq: squared scaling objective} is replaced with the following:
\begin{align}
    ~& \min_{\matr{x}} \sum_{c \in C} \bigg| t_{od}^c - \sum_{e \in E_c} x_e t_e^{flow} \bigg|
\end{align}


\subsection{AMoD Simulation Framework}

The impact of the scaling methods is tested in an agent-based simulation framework FleetPy \cite{engelhardtFleetPyMoDularOpenSource2022}. The framework consists of three main types of agents: customers, a fixed fleet of vehicles $V$, and an AMoD operator. The AMoD operator is responsible for assigning vehicles to dynamically added customer requests. The same historic data used for scaling the network travel times are used for generating the AMoD customer requests. Each customer $r$ requests a ride from the origin location $o_r$ to the destination location $d_r$ at time $t_r$. Let $d^r_{od}$ represent the travel distance from $o_r$ to $d_r$. The AMoD operator tries to maximize the monetary profit for the whole simulation period, which consists of the base fare charged per served customer ($\zeta$), the variable fare charged per unit distance with a customer on board ($f^D$), variable operational cost per unit distance per vehicle ($c^D$) and fixed daily vehicle cost per vehicle $c^F$. Thus, the profit for the entire simulation period is given as:
\begin{equation}
\label{eq: overall objective}
    \sum_{r\in R_s}(\zeta + f^D \cdot d^r_{od})  - \sum_{v \in V} (c^F \cdot n_{days} + c^D \cdot d_v)
\end{equation}
where $R_s$ is the set of all customers served and $n_{days}$ is the number of evaluation days. $d_v$ is the total distance driven by vehicle $v$.

The current paper uses the method of \cite{syedDensityBasedDistribution2021} for the assignment of vehicles to customers: the AMoD operator accumulates the requests into batches using the period $\Delta T_{batch}$ followed by solving an optimization problem that maximizes Eq.~\ref{eq: overall objective} for the current batch. It also includes a hard constraint such that the customers must be picked up within a maximum waiting period $\Delta T_{max}$ of $t_r$. Within the assignment problem, the en-route vehicles are considered from the time and location of their availability. Any customer that is not assigned to a vehicle is removed from the future customer batches. The paper does not consider re-optimization of the assignment problem; once a vehicle is assigned to a customer its path remains fixed. For simplicity, the paper does not consider pooling multiple requests into a single ride. The simulation framework additionally uses a repositioning approach where the excess vehicles are equally distributed among regions after a period of $\Delta T_r$.

\section{Case Study}

\subsection{Experimental Setup and Trips Data}

\begin{figure}[tbp]
  \centering
    \includegraphics[width=1\linewidth]{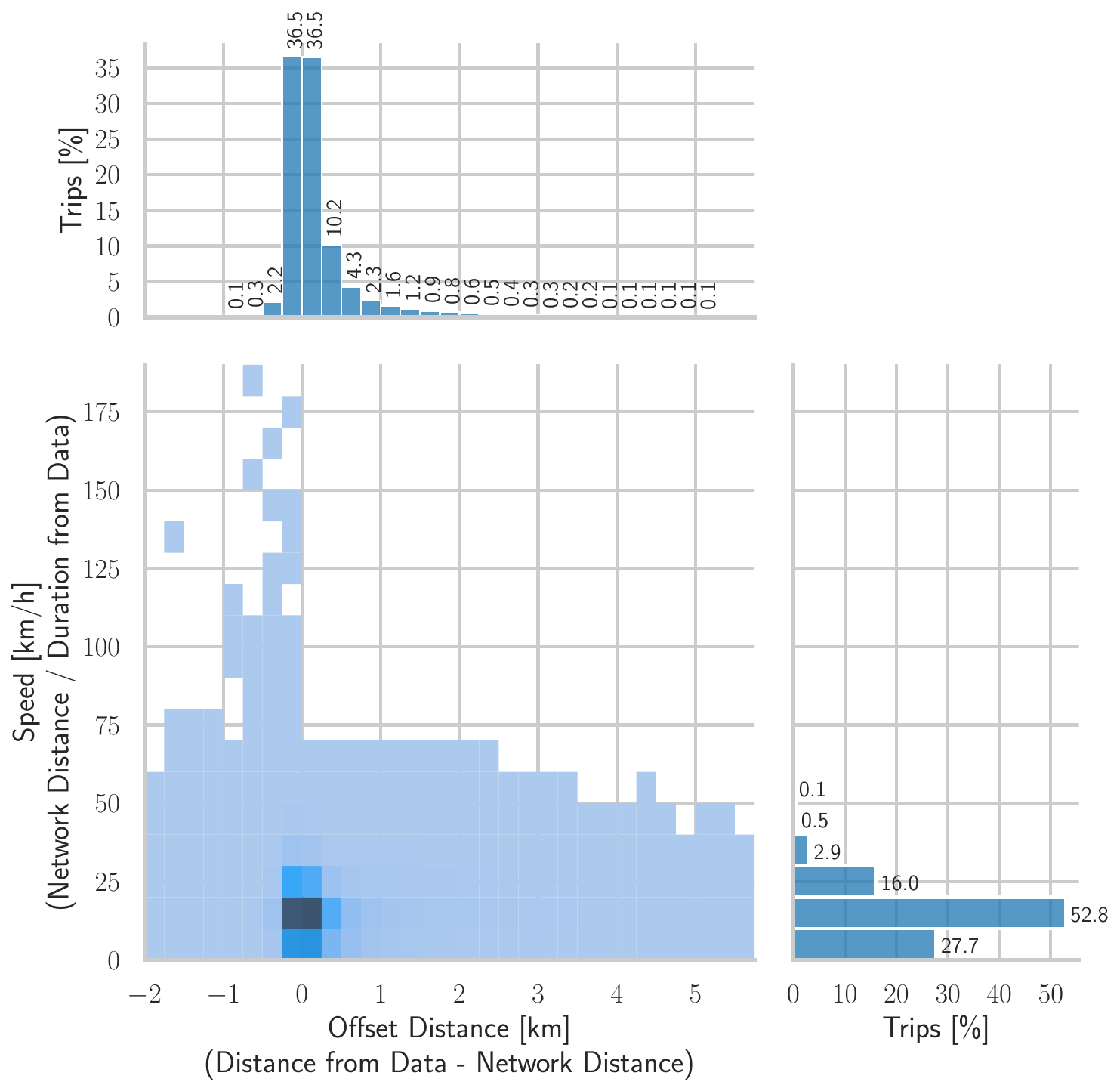}
  \caption{The difference between the trip distances as available from the data and calculated from the OSM network (the shortest path distances using trip origins and destinations). The speed on the y-axis shows the speed as calculated using the shortest path distance from OSM and the recorded travel duration from the data. The data refers to Manhattan trips of NYC Taxi data from 6 to 7 June 2016.}
  \label{fig: distance offset}
\end{figure}

The widely used open-source New York City (NYC) data is used to study the impact of scaling methods on the AMoD simulations. The NYC data from 6 to 7 June 2016 is used for this purpose. The NYC data from 2016 provides the exact coordinates of trip origins and destinations. The data also records the trip's total travel time and distance using which the possible false trips are removed that have speeds less than 1~mph or more than 55~mph. Since most of the trips are within Manhattan, only the trips that start and end in Manhattan are considered. 

Next, since both SSM and ASM methods require trip trajectories for scaling, the paper suggests that the trajectories of the shortest-distance paths calculated using OSM network could be used for this purpose. As shown in Fig.~\ref{fig: distance offset}, for 73\% of the trips, the difference between the distance of the shortest-distance path from the OSM network and the recorded trip distance is less than 250~m; for 92.3\% of the trips, it is less than 1~km. Furthermore, to check if the calculated trajectories also have a meaningful distance in comparison to the recorded travel times, the speed of the trips is calculated using OSM distances and recorded trip duration. The vast majority of the trips are within a reasonable speed range; a very small number of trips have speeds higher than 55~mph (88.5~km/h) for shortest-path trajectories which are also removed. These statistics indicate that the shortest-path routes are very close to the original routes taken by the trips.

For the AMoD simulation, the objective function variables $(\zeta, f^D, c^D, c^F)$ are set as (0.5\$/served customer, 0.5\$/km, 0.25\$/km, 25\$/vehicle per day). $\Delta T_{batch}$, $\Delta T_{max}$ and $\Delta T_r$ are set to 30~seconds, 6~minutes and 30~minutes, respectively. For the scaling methods, $\Delta T_{scale}$ of 30~minutes is used.

\subsection{Results}

\begin{figure}[tbp]
  \centering
    \includegraphics[width=1\linewidth]{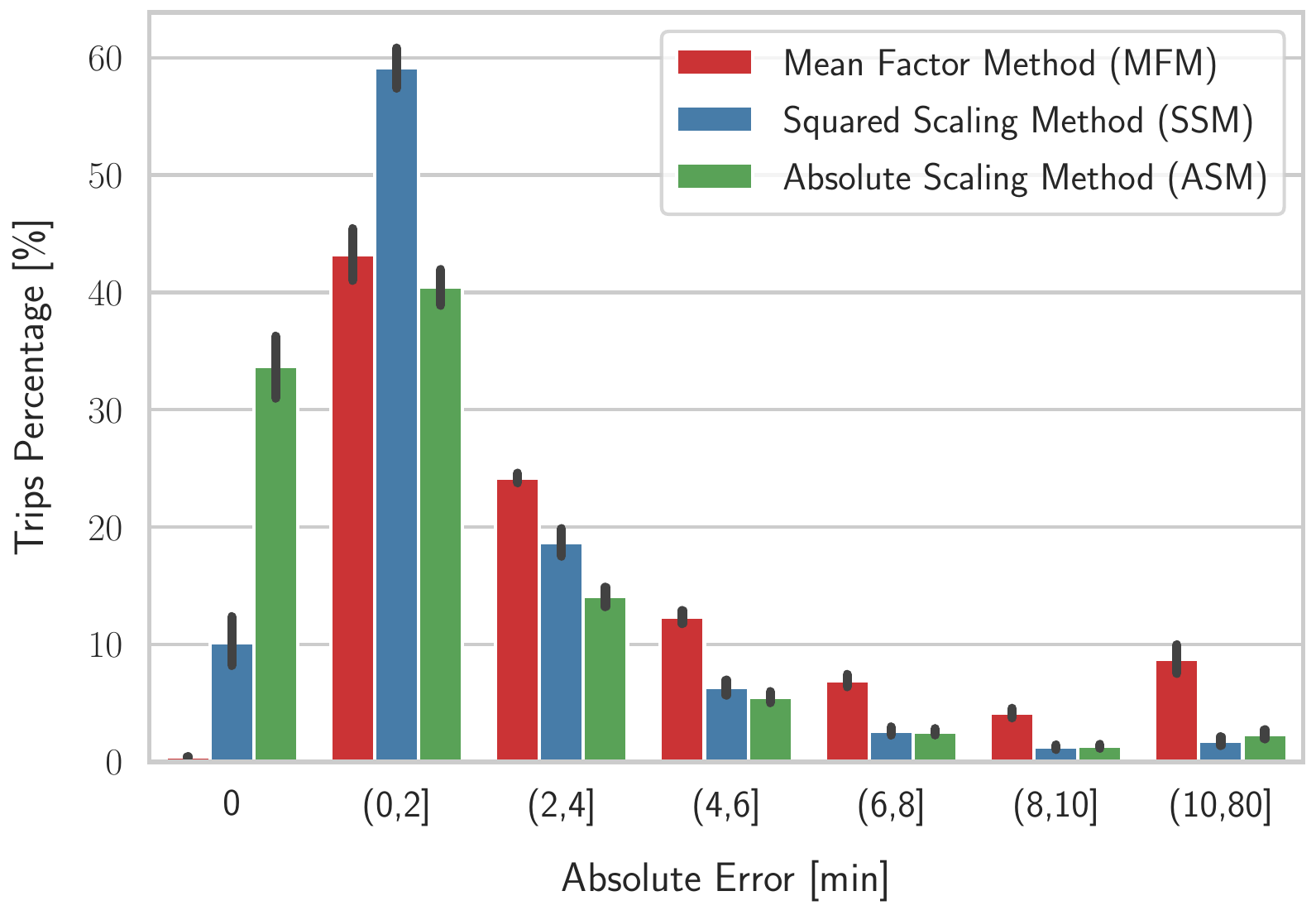}
    \caption{Percentages of trips in different ranges of absolute error. An absolute error less than 1 second is marked a 0.}
    \label{fig: manhattan scaling time range}
\end{figure}

\begin{table}[tbp]
\centering
\begin{tabular}{|l|ccccc|c|}
\hline
\textbf{} & \multicolumn{5}{c|}{Percentile (in minutes)} & \multicolumn{1}{l|}{\textbf{}} \\ \hline
\begin{tabular}[c]{@{}l@{}}Scaling\\ Method\end{tabular} & \multicolumn{1}{c|}{5\%} & \multicolumn{1}{c|}{25\%} & \multicolumn{1}{c|}{50\%} & \multicolumn{1}{c|}{75\%} & 95\% & Maximum \\ \hline
MFM & \multicolumn{1}{c|}{0.23} & \multicolumn{1}{c|}{1.22} & \multicolumn{1}{c|}{2.74} & \multicolumn{1}{c|}{5.26} & 11.99 & 71.14 \\ \hline
SSM & \multicolumn{1}{c|}{0.04} & \multicolumn{1}{c|}{0.49} & \multicolumn{1}{c|}{1.23} & \multicolumn{1}{c|}{2.47} & 5.86 & 59.66 \\ \hline
ASM & \multicolumn{1}{c|}{0} & \multicolumn{1}{c|}{0.03} & \multicolumn{1}{c|}{0.76} & \multicolumn{1}{c|}{2.16} & 6.45 & 69.66 \\ \hline
\end{tabular}
\caption{Percentiles of absolute errors in trip travel times when calculated using scaled city network.}
\label{tab: absolute error for Manhattan network}
\end{table}

\begin{figure*}[tb]
  \centering
  \begin{subfigure}[t]{0.24\linewidth}
    \includegraphics[width=\linewidth]{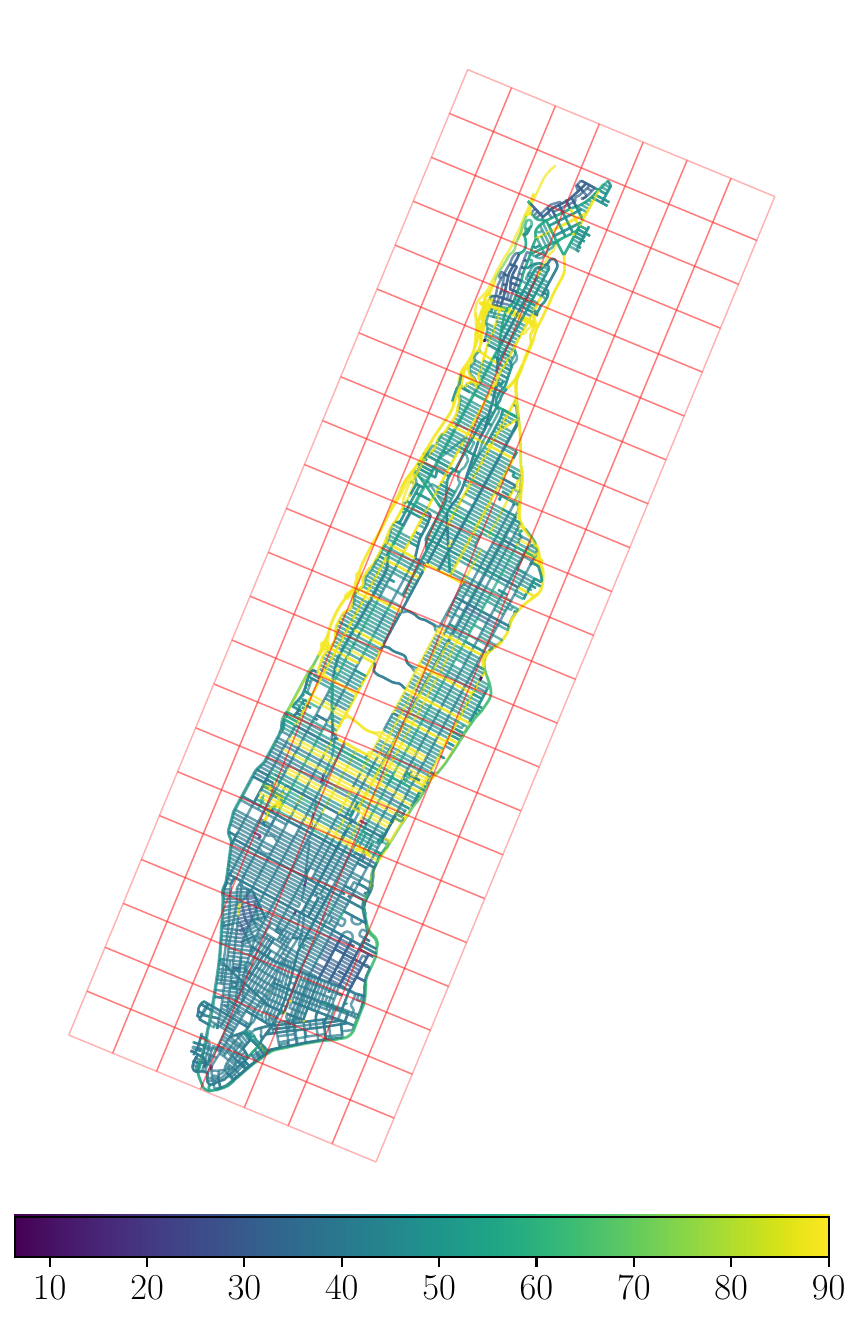}
    \caption{Free-flow speed [km/h]}
    \label{fig: manhattan regional scaling (a)}
  \end{subfigure}
\begin{subfigure}[t]{0.24\linewidth}
    \includegraphics[width=\linewidth]{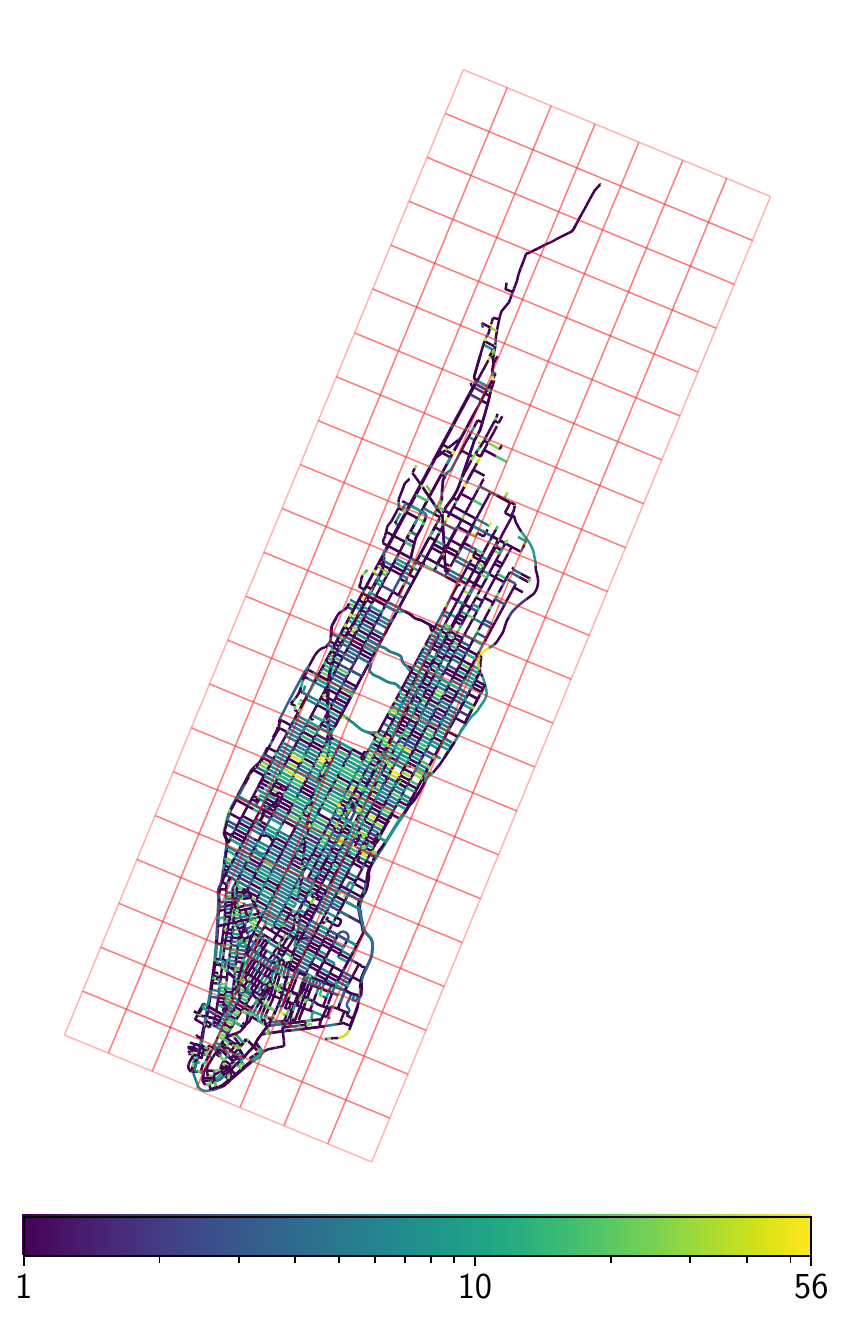}
    \caption{Scaling factors (shortest-distance routes)}
    \label{fig: manhattan regional scaling (b)}
  \end{subfigure}
    \begin{subfigure}[t]{0.24\linewidth}
    \includegraphics[width=\linewidth]{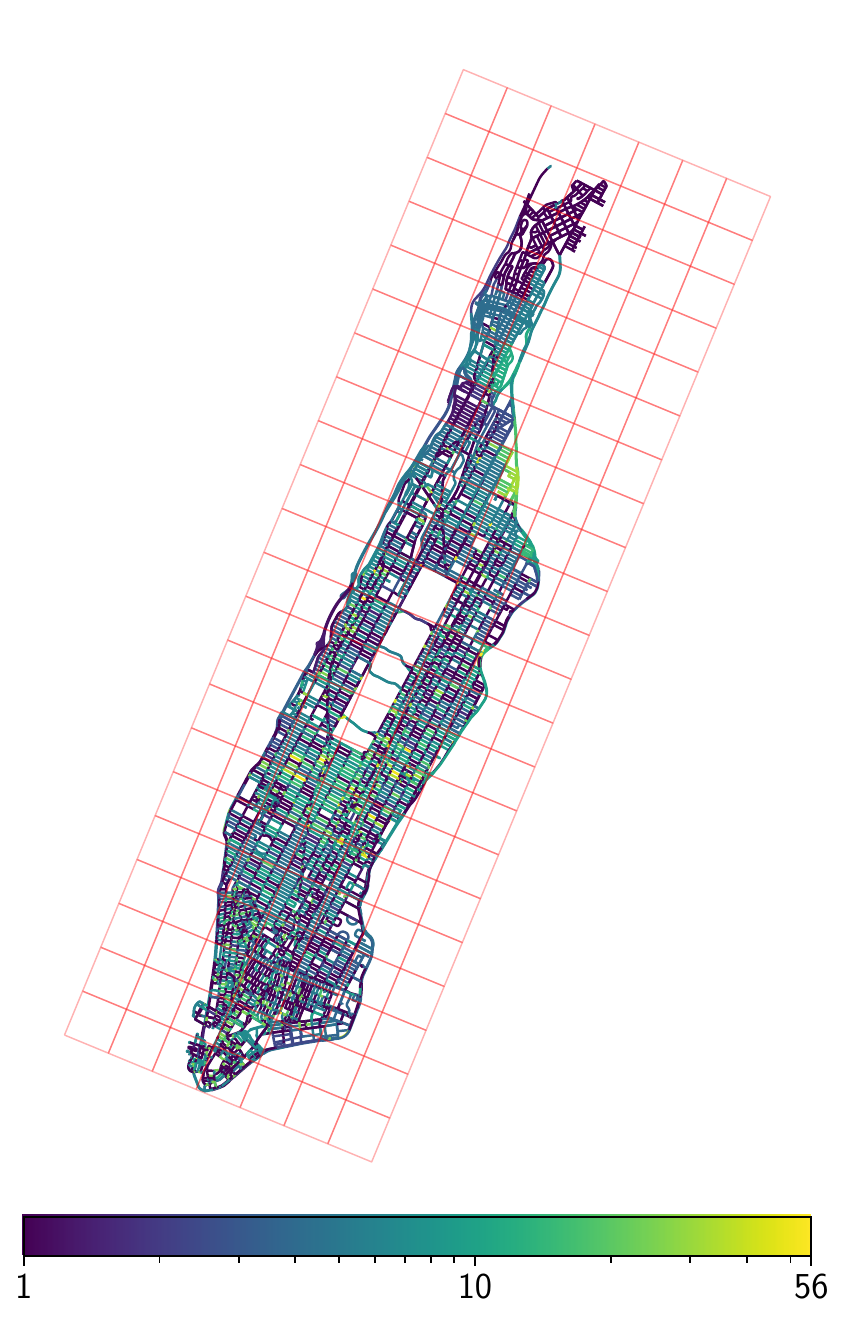}
    \caption{Scaling factors}
    \label{fig: manhattan regional scaling (c)}
  \end{subfigure}
    \begin{subfigure}[t]{0.24\linewidth}
    \includegraphics[width=\linewidth]{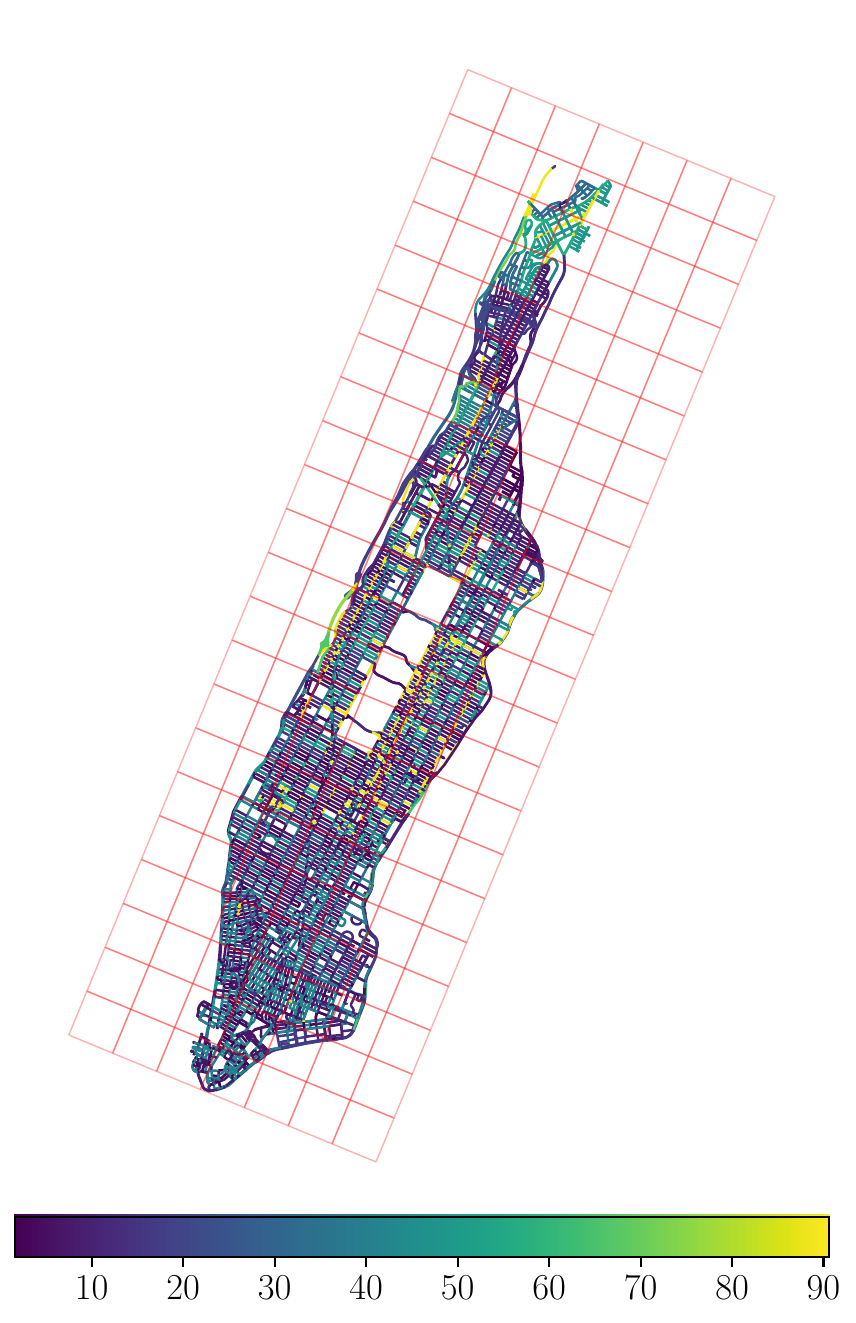}
    \caption{Scaled speeds [km/h]}
    \label{fig: manhattan regional scaling (d)}
  \end{subfigure}
  \caption{An example of scaling factors and speeds obtained using cell size $\Delta s_{cell}$ of 1~km and ASM method. The data used is for the period from 9~am to 9:30~am on 6-6-2016.}
  \label{fig: manhattan regional scaling}
\end{figure*}

Fig.~\ref{fig: manhattan scaling time range} shows the absolute error between the recorded travel times in the data and the travel times of the corresponding trips obtained using the scaled network. Table~\ref{tab: absolute error for Manhattan network} shows the percentiles of the absolute error. First, MFM does not produce a significant proportion of trips with exact accuracy (an absolute error of less than 1 second). Contrarily, due to the link-based scaling, both SSM and ASM reproduced exact travel times for higher a proportion of trips. It also demonstrates a major difference between SSM and ASM: because the SSM uses the square of the scaling error as the objective function, it prioritizes the minimization of high-magnitude errors. Thus, SSM has a lower 95\% percentile and maximum absolute error than ASM as shown in Table~\ref{tab: absolute error for Manhattan network}. Similarly, it has a slightly lower number of trips than ASM with a higher range of absolute error between 10 and 80 minutes. In contrast, the ASM uses the absolute value of the scaling error as the objective function, and thus, it prioritizes the minimization of scaling error for a higher number of trips. Figure~\ref{fig: manhattan scaling time range} also demonstrates this behavior; ASM has a much higher proportion of trips in lower ranges of absolute error. In fact, almost 33.7\% of the trips have no scaling errors with ASM compared to 10.2\% for SSM and only 0.42\% for MFM. In terms of percentages, 50\% of the trips using MFM method have an absolute error of less than 2.74 minutes while using the SSM and ASM methods it reduces to 1.23~minutes and 45.6~seconds, respectively.
Thus, the spatio-temporal scaling would produce significantly more realistic travel times between corresponding origin and destination locations in AMoD simulations than MFM.

Fig.~\ref{fig: manhattan regional scaling} illustrates an example of the ASM scaling for the rush hour (9~am to 9:30~am) of Monday morning. First, as shown in Fig.~\ref{fig: manhattan regional scaling (c)}, the scaling factor varies for different regions and links in the network. Second, as shown in Fig.~\ref{fig: manhattan regional scaling (b)}, the shortest-path routes do not necessarily include all of the network edges; rather, the areas with the highest number of origins or destinations (e.g., central Manhattan for the rush hour) provide the highest level of information for the scaling method. This is because the scaling is done in intervals of $\Delta T_{scale}$ (30 minutes) and only the trips within $\Delta T_{scale}$ are used for scaling, and thus, some regions may contain very few edges from $E_u$ while others may not have any edge from $E_u$ at all (refer to the cells in the north in Fig.~\ref{fig: manhattan regional scaling (b)}). Therefore, the scaling accuracy of a region also depends on the number of historical trips and shortest path edge samples available for a particular region. This is apparent by comparing the regions in Fig.~\ref{fig: manhattan regional scaling (b)} that have a small number of edge samples to the edges in Fig.~\ref{fig: manhattan regional scaling (c)} and the resulting scaled speed in Fig.~\ref{fig: manhattan regional scaling (d)}. However, this also depends on the time of the day as the pattern of trip origins and destinations change from morning to evening and during night hours. Nevertheless, both ASM and SSM methods compensate for the absence of the data by assigning the average scaling factor of a cell to the links without a travel time sample. This assumes that the traffic situation within a small cell should be similar for all the links in that specific cell.

\begin{figure*}[tbp]
  \centering
    \includegraphics[width=0.93\linewidth]{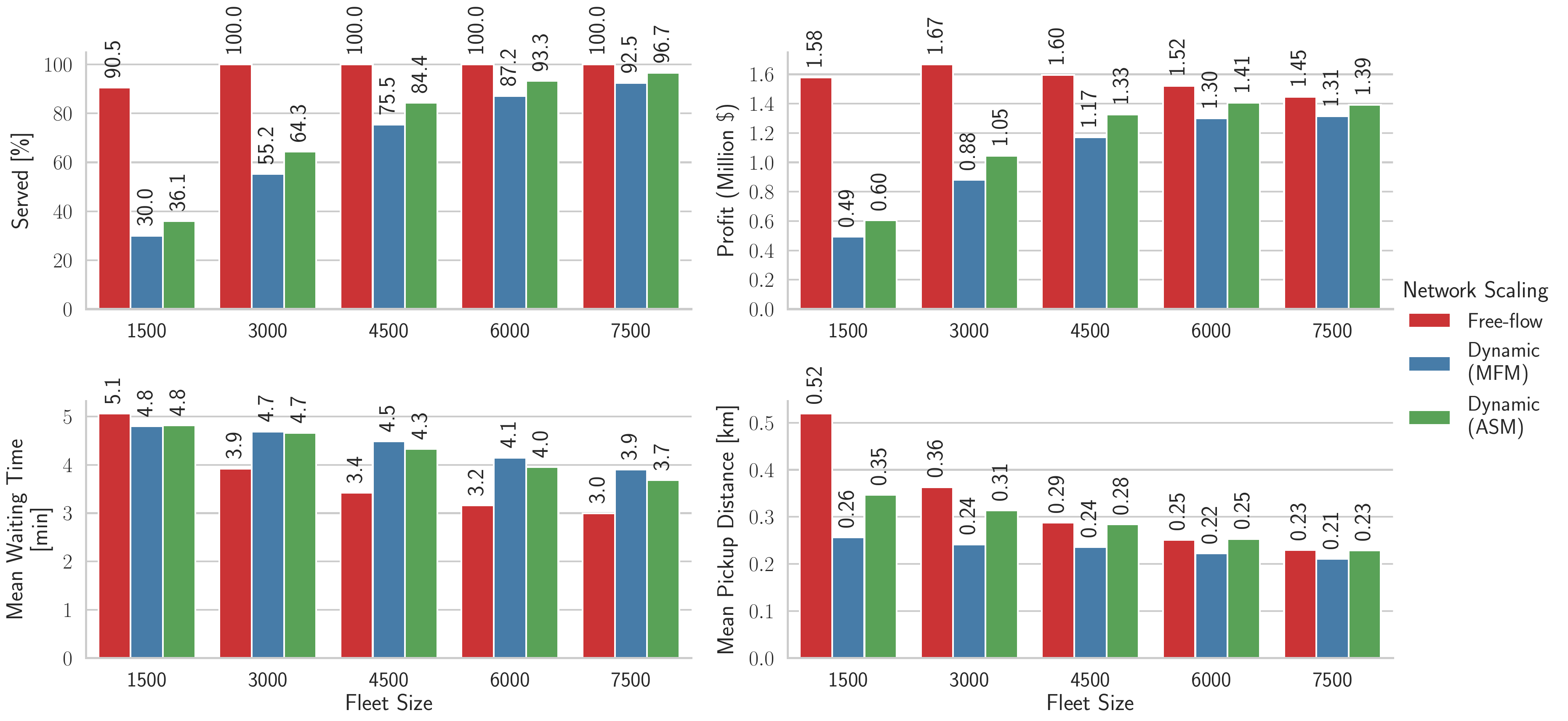}
    \caption{Impact of the scaling method on AMoD simulation results. The free-flow represents the speeds obtained from OSM maps which highly overestimates the performance while the MFM method underestimates it.}
    \label{fig: simulation performance}
\end{figure*}

Finally, Fig.~\ref{fig: simulation performance} compares the impact of scaling methods on the AMoD simulations. First, the free-flow speeds significantly overestimate the AMoD performance: even a small fleet of 1500 vehicles is able to serve almost 90.5\% of the customers and generate a very high monetary profit of \$1.58 million. Due to the high speed of the network, the AMoD vehicles are able to quickly serve the customers as well as pick up the customers from far-off distances with short waiting times, as shown in Fig.~\ref{fig: simulation performance}. 

With free-flow speeds and larger fleet sizes, all of the customer requests can be served by the AMoD service; however, the additional maintenance cost reduces the overall profit. On the contrary, when more realistic travel times are used the AMoD simulation results significantly vary from the free-flow speed. In general, the AMoD performance is reduced when more realistic travel times are used. In comparison to the free-flow, a fleet of 1500 vehicles only serves 30\% of the requests and produces a profit of only \$0.49 million. Even a significantly high fleet of 7500 vehicles only serves 92.5\% of the requests. However, the simulation results show that in contrast to the overestimation of free-flow, the MFM underestimates the AMoD performance; on average, with ASM scaling the AMoD simulation serves 6.88\% more customers and shows a relative increase of 14\% in profit compared to MFM scaling. The calculation of scaling factors in MFM implicitly gives a higher weightage to the areas of higher customer demand which most of the time also happens to be the areas with lower travel speeds. Since the whole network is then multiplied by the common factor, this leads to a relatively slower network with MFM method.

\section{Conclusion and Limitations}

This paper presented a spatio-temporal scaling method for the AMoD simulation. Many times the AMoD simulations use real trips data from Taxi or MoD services to study the impacts of various control strategies. The paper suggests that a more realistic and reliable AMoD simulation can be produced by spatio-temporal scaling of the travel times using the real trip data already used for generating customer requests. This makes sure that the travel times are as close as possible to the original trips. The approach was tested for an AMoD scenario in Manhattan using NYC Taxi data. The results show that the approach produces significantly high accuracy compared to other scaling methods commonly used for AMoD simulation.

The introduced methods are specifically formulated for the AMoD simulations. The work assumes a simplistic city network obtained from OSM without considering detailed aspects of a real network such as different lane speeds and traffic lights. These are only implicitly taken into account by reducing the errors against historic trip data. It was also assumed that the links with missing data can be scaled using the average speed in the area. However, this basic assumption may not necessarily be true for all networks. Additionally, Manhattan is a relatively small area with a significantly high number of trips. Other real data used for AMoD simulations may not have such a high number of trips available. Thus, the results may vary when the method is applied to other data sets.

In the future, a regularization factor can be introduced that smooths out the variation of link speeds within and across regions. Instead of using regular grids, a better clustering of trips can be focused. Additionally, the approach can be combined with other data sources for improved scaling of the network for missing links.

\bibliographystyle{IEEEtran}
\bibliography{IEEEabrv,library.bib}

\end{document}